\documentclass[amsmath,amssymb,twocolumn,nofootinbib]{revtex4-1}

\usepackage[utf8]{inputenc}
\usepackage{times,bbm}
\usepackage{amsfonts}
\usepackage{xcolor}
\usepackage{mathtools}

\usepackage{graphicx}
\usepackage{float}
\usepackage{csquotes}

\usepackage{hyperref}
\hypersetup{
	breaklinks,
	colorlinks,
	linkcolor=gray,
	citecolor=gray,
	urlcolor=gray,
}

\usepackage{dsfont}
\newcommand{\Id}{\mathds{1}}
\newcommand{\Tr}{\text{Tr}}
\newcommand{\ket}[1]{\left \vert #1 \right \rangle}
\newcommand{\bra}[1]{\left \langle #1 \right \vert}

\newcommand{\ZZ}{\mathcal{Z}}

\definecolor{jens}{rgb}{.2,0.7,.9}
\definecolor{cadmiumgreen}{rgb}{0.0, 0.42, 0.24}

\definecolor{marek}{rgb}{.5,.5,.2}

\begin{document}

\title{Are many-body localized systems stable in the presence of a small bath?}

\author{Marcel Goihl, Jens Eisert and Christian Krumnow}
\affiliation{Dahlem Center for Complex Quantum Systems, Freie Universit{\"a}t Berlin, 14195 Berlin, Germany}

\date{\today}

\begin{abstract}
  When pushed out of equilibrium, generic interacting quantum systems
  equilibrate locally and are expected to 
  evolve towards a locally thermal description despite their unitary time evolution. Systems in which disorder 
  competes with interactions and transport can
  violate this expectation by exhibiting many-body localization. 
  The strength of the disorder with respect to the other two parameters drives a transition from a thermalizing system towards a 
  non-thermalizing one. The existence of this transition is well established both in experimental and numerical studies for finite systems. 
  However, the stability of many-body localization in the thermodynamic limit is largely unclear. With increasing system size, a generic 
  disordered system will contain with high probability areas of low disorder
  variation. If large and frequent enough, those areas constitute ergodic
  grains which can hybridize and thus compete with localization.
  While the details of this process are not yet settled, it is conceivable that
  if such regions
  appear sufficiently often, they might be powerful enough to restore thermalization.
  We set out to shed light on this problem by constructing potential landscapes
  with low disorder regions and numerically investigating their localization behavior
  in the Heisenberg model.
  Our findings suggest that 
  many-body localization may be much more stable than anticipated in other recent theoretical works.
\end{abstract}
\maketitle

\section{Introduction}

One of the long-standing puzzles of physics is how the postulates of quantum statistical mechanics 
and thermodynamics can be made compatible with the unitary time evolution of
quantum systems.
It is increasingly becoming clear that generic interacting
quantum systems -- once pushed out of equilibrium -- are expected to dynamically evolve towards a
locally thermal description again \cite{rigol08,polkovnikov11,eisert15,christian_review,langen15}. This constitutes
an interesting state of affairs, since it reconciles the apparent contradiction
between the description of 
time-evolving states and of equilibrium ensembles. Such an  interpretive scheme also provides
a picture in which interacting systems can be described by only a small number of 
parameters for almost all time intervals, thus avoiding the curse of dimensionality.
A few exceptions are known to exist but these are fine-tuned 
\emph{integrable} systems featuring local constants of motion that 
prohibit a general description in terms of thermal ensembles. 

At the turn of the millennium, a new
class of quasi-integrable systems emerged which are fail to thermalize over a wide range of
parameters. As this effect is caused by the subtle interplay of transport,
interactions and disorder, it has been dubbed
\emph{many-body localization (MBL)} \cite{gornyi05,basko06,oganesyan07}. 
A many-body localized system does not exhibit transport of 
particle-like quantities and therefore can be effectively described by an extensive set of quasi-locally
conserved constants of motion (qLCOMs) \cite{serbyn13,huse14}. This leads to local memory of particle configurations
\cite{schreiber15}. Note that unlike the non-interacting Anderson insulator, 
systems exhibiting MBL may well transport information-like quantities such as
correlations between particles. 
This is reflected e.g. in the logarithmic growth of the entanglement entropy in time 
\cite{znidnaric08,bardarson12}.

While many of the properties ascribed to MBL have been observed either experimentally or numerically 
in finite systems, the question of the 
stability of MBL in the thermodynamic limit is as of yet unresolved.
The expected leading sources of instability are rare regions of low disorder that might conceivably
thermalize the rest of the system \cite{ponte17,deroeck17,luitz17,thiery18}. 
Questions regarding the effect of local regions that are 
localized in the otherwise ergodic phase or ergodic regions in the otherwise localized
phase have been investigated in the field of \emph{Griffiths effects}
\cite{nandkishore14,gopalakrishnan15,agarwal15,luitz172,nandkishore17,gopalakrishnan19}
(for a review, see Ref.~\cite{agarwal17}). The question of whether a closed
MBL system is stable to uncharacteristic disorder potentials is hence part of
this sub-field.

In this work, we deliberately construct 
potential landscapes with regions of improbably low disorder and study their
influence on a localized chain surrounding them. 
A previous study has reported that a constant size thermal region is able to thermalize a localized
chain coupled to it independent of its size if the constants of motion do not decay sufficiently strongly  
\cite{luitz17}. The study has been conducted within an effective description of
MBL \cite{serbyn13,huse14}, one in which
the localized part has been modelled by constants of motion and the bath by a suitable 
random matrix. The postulated coupling between bath and localized part is characterized by a decay length which is related to the localization length of the constants of motion. 

Here, instead we use its real space equivalent; the disordered Heisenberg chain. Employing exact diagonalization, we analyze the statistics 
of local expectation values of the eigenstates which allow us to draw conclusions 
about the locality of the constants of motion of the system. Our results are not entirely
compatible with those of Ref.\ \cite{luitz17}, as we find MBL is not compromised by the presence of low disorder 
regions of constant size. When the size of the region is allowed to scale with system size however, we find that
localization vanishes 
when extrapolating our results to large systems.

\section{Setting}

\subsection{Hamiltonian model}
\begin{figure*}
        \includegraphics[width=\textwidth]{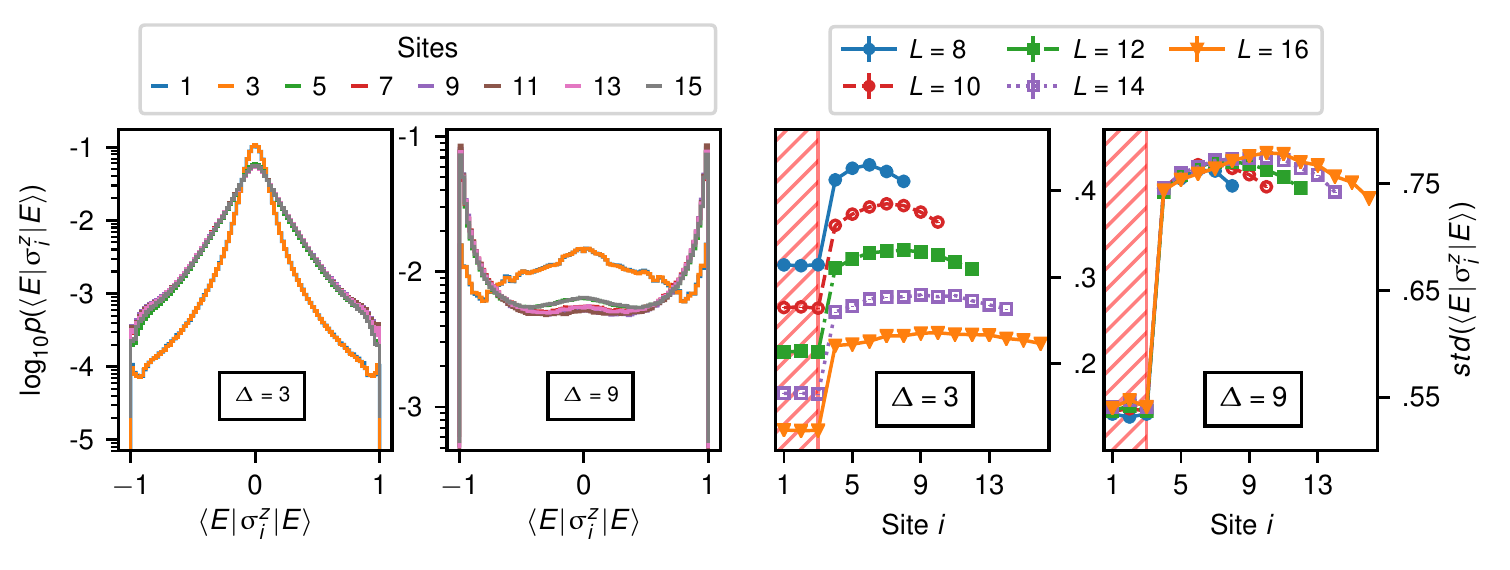}

  \caption{Local expectation values of the eigenstates of the model
	with three thermal sites for weak ($\Delta=3$) and strong disorder ($\Delta=9$). Left:
	Histograms of $\bra{E} \sigma^z_i \ket{E}$ 
  in the Heisenberg model on $L=16$ sites for $i \in \{1,3,\ldots,15\}$
  encoded by color. 
  Each histogram is an average over all eigenstates in the 
  zero magnetization sector and 2000 realizations.
	Right:
  Standard deviation of the histograms of
  $\bra{E} \sigma^z_i \ket{E}$ 
  in the Heisenberg model for different system sizes.
 	Red hatches show the thermal part of the system.	
  Each data
  point is an average over at least 2000 realizations.}
  \label{full_3}
\end{figure*}

\begin{figure}
  \includegraphics[width=0.5\textwidth]{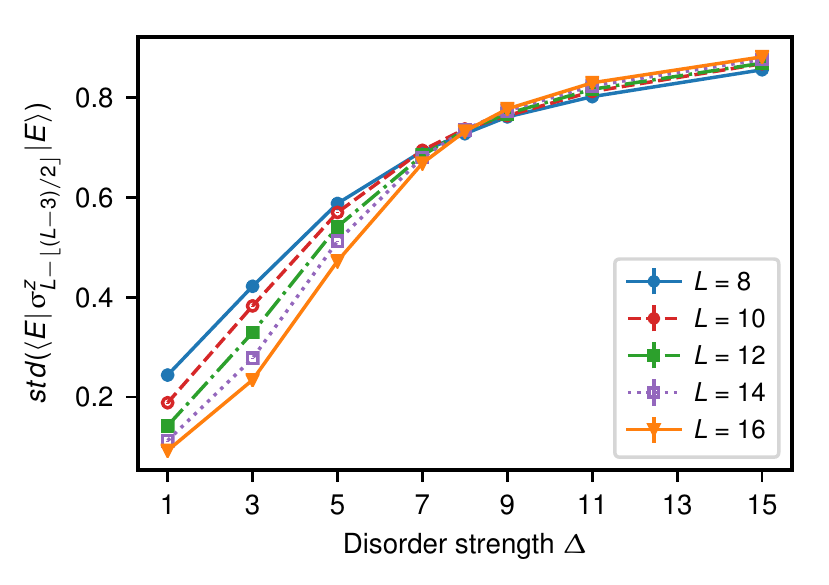}
  \caption{
  Standard deviation of the histograms of 
	$\bra{E} \sigma^z_i \ket{E}$ of the center site
	in the disordered part positioned 
	at $i =L- \lfloor (L-3)/2 \rfloor$
  in the Heisenberg model with 3 thermal sites for different 
  system sizes. Each data point is an average over 
  at least 2000 realizations.}
  \label{std_3}
\end{figure}
We consider the ``drosophila'' of MBL, the disordered 
spin-$1/2$-Heisenberg chain. 
In order to study the effect of small subregions of low disorder we consider systems where the potential on the sites $1,\dots,s$ is close to zero. For a system of $L$ sites our Hamiltonian then reads
\begin{align}
\label{eq:Heis}     
  H &= \sum\limits_{i=1}^L \left(\sigma^{x}_i\sigma^{x}_{i+1}+
  \sigma^{y}_i\sigma^{y}_{i+1}+ \sigma_i^{z} \sigma^{z}_{i+1} \right)\nonumber\\
  &\quad+\sum\limits_{i=1}^s \epsilon\, h_i\sigma^{z}_i+\sum\limits_{i=s+1}^L \Delta\, h_i\sigma^{z}_i\,,
\end{align}
where the $h_i$ are drawn uniformly and independently from the interval $[-1,1]$,
$\epsilon$ and $\Delta$ denote the disorder strength in the low and high
disorder region and $\sigma^a_i$ is the Pauli-$a$ operator
acting on the $i$-th site. 
The fully disordered model (without ergodic subregion) is expected to undergo a localization 
transition at a critical disorder strength $\Delta_c \approx 7$ 
(note that since we use Pauli instead
of spin operators, the transition is shifted with respect to other literature).
Moreover, we use periodic boundary conditions and work in the
zero magnetization sector.
We set $\epsilon = 10^{-6}$ independent of the disorder
strength $\Delta$. This creates the situation that the local flatness of the potential
may thermalize the subsystem, which is expected to compete
with localization effects in the remaining system. In the following, we will hence refer to the first $s$
sites as the thermal sites and refer to the remaining $L-s$ sites as the disordered sites or part of the system. 
We consider two different scenarios:
In the first, the number of thermal sites is independent of the system size. 
In the second setting, we take a fixed fraction of the full size to be thermal.

\subsection{Distribution of disorder}

Within this model, we investigate below the effect of a local cluster with
uncharacteristically small disorder on its environment. The formation of such a cluster
should be understood as a rare instance of a generic MBL model and is specifically relevant when considering
the stability of an interacting localized phase in the thermodynamic limit,
where the effect of these rare regions is not fully understood yet. For large
systems, the probability of having small thermal regions increases and poses
the question if these are able to compromise localization of the full chain 
as well. As a prerequisite, these must first thermalize their neighbours 
which can also be checked in small systems and which is precisely 
the context of this work. Note that if one tries to draw conclusions for larger systems 
from our results, these can only hold if the reductions of the eigenstates
resemble the eigenstates of the small system when probed locally. 
Global quantities such
as energy levels commonly used to detect MBL \cite{oganesyan07,luitz15} will 
likely not be faithfully recovered in our small system and are hence not considered here. 

\section{Measure of localization}
\begin{figure*}
        \includegraphics[width=\textwidth]{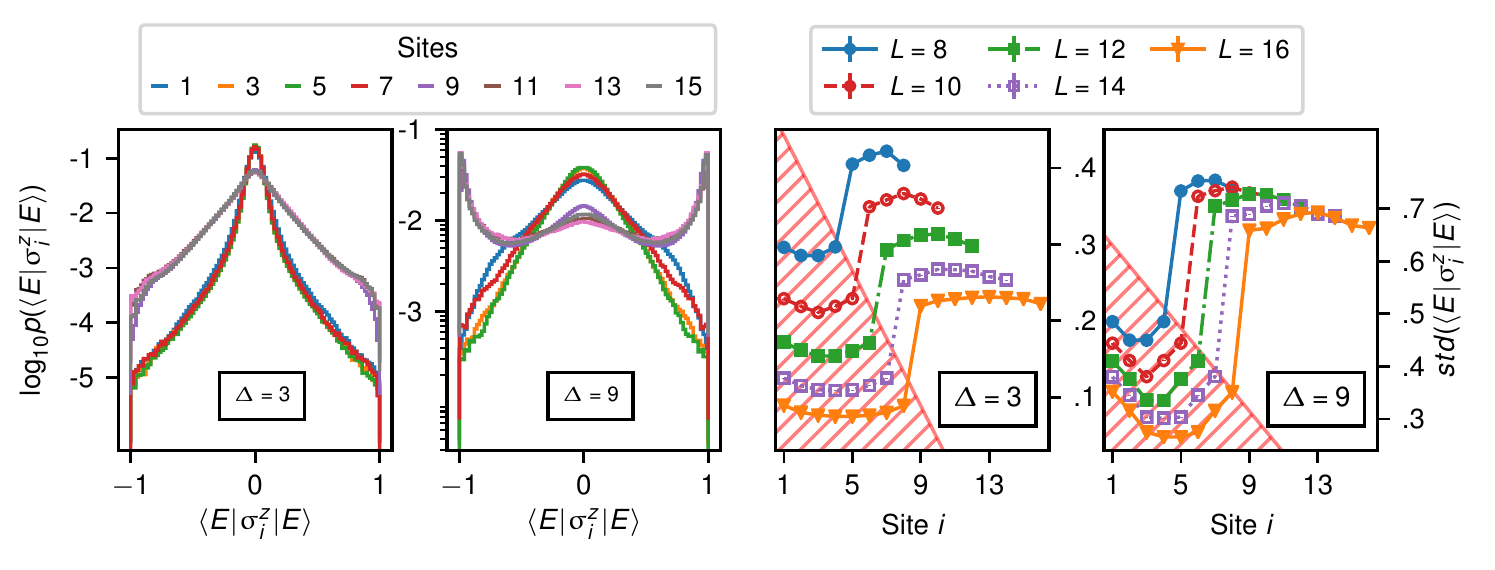}
  \caption{Local expectation values of the eigenstates of the model
	with $L/2$ thermal sites for weak ($\Delta=3$) and strong ($\Delta=9$) disorder. Left:
	Histograms of $\bra{E} \sigma^z_i \ket{E}$ 
  in the Heisenberg model on $L=16$ sites for $i \in \{1,3,\ldots,15\}$
  encoded by color. 
  Each histogram is an average over all eigenstates in the 
  zero magnetization sector and 2000 realizations.
	Right:
  Standard deviation of the histograms of
  $\bra{E} \sigma^z_i \ket{E}$ 
  in the Heisenberg model for different system sizes.
 	The red hatched part of each curve indicates the thermal region of the system.	
  Each data
  point is an average over at least 2000 realizations.}
  \label{full_l2}
\end{figure*}
The proposed \emph{effective integrability} of MBL comes about due to the
presence of extensively many quasi-local conserved operators $\ZZ_i$
\cite{serbyn13,huse14}. 
This is seen most readily in the infinite disorder limit 
($\Delta \rightarrow \infty$) where these are given by the local fields
$\ZZ_i = \sigma^z_i$ and hence act on a single site only. Moreover, they allow
to label the eigenstates by the occupation of the $\ZZ_i$. When one moves away from
the infinite disorder case, energy space and real space no longer coincide
and the above identity becomes more intricate. 
The Hamiltonian formulated in real space is then
diagonalized by a unitary $U_D$ which relates energy and real
space. The real space representation of the $\ZZ_i$ is given by a
decomposition in which all possible Pauli operator combinations can appear
\begin{equation}
  U_D \ZZ_i U_D^\dagger = \alpha_i \sigma^z_i + \sum_{\mu,\nu \in C} \beta_i(\mu,\nu)
  \Sigma^z(\mu) \Sigma^x(\nu)\,,\label{eq:lcomexpansion}
\end{equation}
where $\Sigma^z,\Sigma^x$ are Pauli words consisting of local Pauli-z and
Pauli-x operators respectively and $\mu,\nu \in \{0,1\}^L$ indicate the position 
of the Pauli operators in the chain, e.g. $\Sigma^a(\mu) = \otimes_i
(\sigma^a)^{\mu_i}$. 
We deliberately singled out the weight on $\sigma^z_i$ corresponding to the infinite disorder
conserved operator.
All remaining weights are subsumed by the index set $C$ which contains
all possible $\mu,\nu$ configurations except the one which would yield
$\Sigma^z(\mu) \Sigma^x(\nu) = \sigma^z_i$.
In practice, $\alpha_i, \beta_i(\mu,\nu)$ can be calculated using the
Hilbert-Schmidt scalar product $\beta_i(\mu,\nu) = \Tr(U_D \ZZ_i U_D^\dagger
\sigma^z(\mu) \sigma^x(\nu))/2^L$. 
This representation might seem cumbersome at first glance,
but also allows one to
understand how a transition between strictly local constants of motion in the
infinite disorder case to non-local constants of motion in the ergodic regime 
can be captured formally in terms of the structure of the constants of motion.
In fact this is nothing but a
formalization of the "dressing" process \cite{serbyn13,huse14}. The above
decomposition is completely general in the sense that the constants 
of motion of any system of qubits can be expanded as
in Eq. (2) and can hence be applied to the localized as well as
to ergodic phase of an MBL system.

The behavior of the conserved operators can be tracked by analyzing the
statistics of $\alpha_i$.
In the localized regime the $\ZZ_i$ are expected to be quasi-local
such that the corresponding weights need to decay strongly for
operators that have support outside of the localization length, an intuition that 
has been confirmed numerically \cite{chandran15,he16,rademaker16,mierzejewski17,thomson18,kulshreshtha18,goihl18,Tarantino}. 
The largest weight will hence still be given by $\alpha_i \sim 1$  and all other
weights should be significantly smaller. In the ergodic regime,
however, $\ZZ_i$ are not local at all causing the weights to smear
out over many different operators and therefore $\alpha_i$ will be 
essentially random and small.

To access the statistics of $\alpha_i$, our main numerical tool is constituted by local
magnetizations of eigenstates, i.e.$\bra{E} \sigma^z_i \ket{E}$ \cite{pal10,luitz17}.
As the eigenstates can be expressed in terms of the projectors $(\Id \pm \ZZ_i)/2$ onto the
(un)occupied sectors of the qLCOM, calculating the expectation value with
$\sigma^z_i$ yields $\pm \alpha_i$ depending on the occupation. 
Therefore, we analyze the histograms over all eigenstates, expecting two 
very distinct regimes.
A localized phase is expected
to exhibit a bimodal distribution peaked at $\pm 1$. This holds whenever
the constants of motion are quasi-local and hence most of the
operator weight is still on the onsite magnetization implying $\alpha_i \sim 1$.

For ergodic systems however, the distribution of the $\alpha_i$ values should feature a
close-to-Gaussian shape with zero mean. Here, the constant of motion will be spread
out over many different operators and therefore the weight on $\sigma^z_i$ 
is essentially random. Note that in principle,
one could also check the other weights $\beta_i(\mu,\nu)$ which certainly yield
further insights into the decomposition of the qLCOMs. There are however
$4^L-1$ many of these and without prior knowledge about which one should be
sampled, this task is computationally infeasible.

\begin{figure}
  \includegraphics[width=0.5\textwidth]{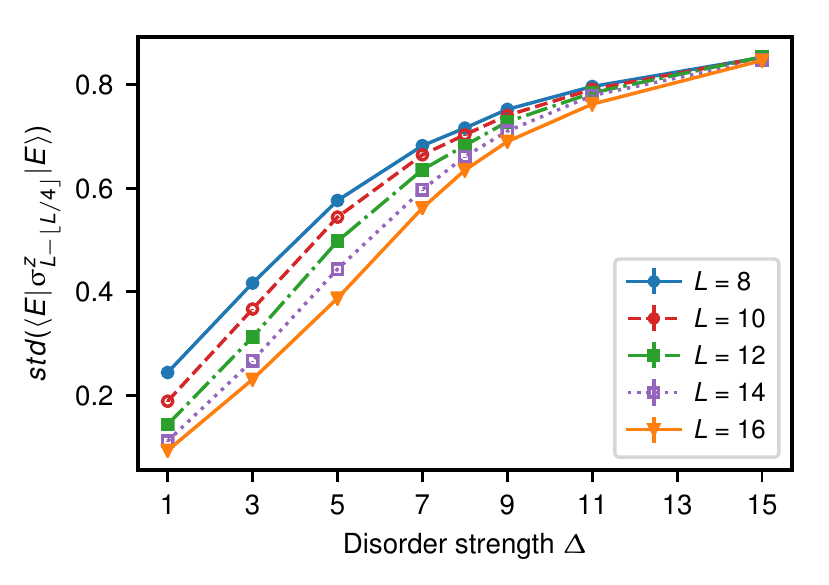}
  \caption{
  Standard deviation of the histograms of 
	$\bra{E} \sigma^z_i \ket{E}$ of the center site
	in the disordered part positioned 
	at $i =L- \lfloor L/4 \rfloor$
  in the Heisenberg model with $L/2$ thermal sites for different 
  system sizes. Each data point is an average over 
  at least 2000 realizations.}
  \label{std_l2}
\end{figure}

\section{Results}
In this section, we show and discuss the obtained results. We worked
with system sizes $L\in \{8,10,12,14,16\}$ and various disorder strengths.
Each point is an average over at least 2000 realizations. Errors have
been calculated either using the standard deviation or
bootstrapping which amounts to resampling from
the obtained data to obtain a distribution of the quantity of interest.
In all plots, the error bars are smaller than the symbols used.
We either set three or $L/2$ sites to be thermal to cover the system size
independent and dependent case. We first discuss the case of the constant
size thermal region.

\subsection{Constant size thermal region}

In the model considered in this section, we set $s=3$, i.e., three sites
are thermal independent of the system size.
The two plots on the left in Fig. \ref{full_3} show the histogram
of the local expectation values of magnetization operators
$\bra{E} \sigma^z_i \ket{E}$ for different sites $i$ encoded in color.
We use a system of size $L=16$. As pointed out above,
this amounts to sampling the weight of the constants of motion on the $\sigma^z_i$
operator and hence is an indirect measure for their locality.
For low disorder $\Delta = 3$, the distributions show a distinct peak at 
zero and feature strongly decaying tails consistent with the predictions for the thermal 
region. When comparing the thermal sites
to the ones that experience the full disorder strength, we find that 
the decay for sites inside the thermal region is stronger -- as one would intuitively expect. 
This implies that the constants of motion are not very dominantly supported on the
respective $\sigma^z_i$ operator but presumably rather spread over all
remaining operators and hence non-local.

The picture changes drastically with increasing disorder. 
For $\Delta = 9$, the distributions originating from sites in the
disordered part of the chain show a distinct bimodality. This shows that the 
$\sigma^z_i$ are very close the exact constants of motion as they inherit their
occupation statistics which is a sign of quasi-locality.
For sites in the thermal region, we find that their distributions feature a
peak around zero and  small peaks in their heavy tails as well.
Since the disorder that these sites experience directly is of order $\epsilon = 10^{-6}$, 
the peaks and the heavy tails should be ascribed to the 
proximity to the localized chain. 

We will now turn to a system size scaling of the observed locality behavior
of the constants of motion.
Since the standard deviation measures
the width of a distribution, it is a genuine measure to compare the
distributions for different system sizes \cite{luitz17}. This is shown
in the two plots on the right in Fig. \ref{full_3},
where we plot the standard deviation for all available
system sizes encoded in color as a function of disorder strength 
in the Heisenberg model with three thermal sites. As a guide to the eye,
the ergodic region is hatched in red. 
For both, weak and strong disorder, we find in Fig. \ref{full_3} 
that the distributions of local expectation values of energy eigenstates are narrower
for sites in the thermal region than in the disordered one.
In the ergodic regime for $\Delta =3$, we see that in both regions the standard deviation 
decreases on all sites with increasing system size.
This is an indicator that the system tends towards thermalization as even
in the disordered part of the system the distributions become more narrow upon increasing the system
size, which is to be expected for $\Delta < \Delta_c$
However, for $\Delta = 9$, 
we find that the standard
deviation shows no systematic shift with increasing system size, instead a
saturation seems to be the most accurate description. This suggests that localization
is not compromised by a constant size thermal region.

To detail this observation, we 
show the standard deviation of a site in the middle of the disordered part of the system at
$i =L- \lfloor (L-3)/2 \rfloor$ as a function of disorder strength in Fig.\,\ref{std_3}.
In the ergodic regime $\Delta < \Delta_c$,
the standard deviation of the middle site decays with increasing system size.
This implies that the constants of motion become less local
when increasing the system size.
Upon increasing the disorder strength, we observe a crossing
behavior of the standard deviation for disorder values higher than the
critical disorder $\Delta>\Delta_c$. This shows that in the localized phase
there is a(n admittedly weak) tendency towards higher standard deviation for increasing system
sizes indicating
that the model is driven towards localization upon increasing system size in
the sense that the constants motion become more and more localized. 
These results indicate, that the presence of a fixed size region of low
disorder acting as ergodic grain does not alter the qualitative behavior of the full system --- depending on the disorder present we find it to be in the ergodic or localized phase without a shift of the transition point compared to the standard disordered Heisenberg chain.

In the next section, we will increase the size of the thermal region with the
system size and carry out the same analysis.

\subsection{Constant fraction thermal region}

Let us now consider the case $s = L/2$ such that the thermal region covers half of the system and
is thus extensive.
The two plots on the left in Fig. \ref{full_l2} show the histogram
of the local expectation values of magnetization operators
$\bra{E} \sigma^z_i \ket{E}$ for different sites $i$ in a system of size $L=16$. 
For low disorder $\Delta = 3$ the distribution essentially has the same shape as for
the three site thermal region with the only exception being that now the eight thermal 
sites show the strong decay and the eight disordered sites decay less strongly.
Again, this implies that the constants of motion are not very dominantly supported on the
respective $\sigma^z_i$ operator.
For $\Delta = 9$ the distributions originating from sites in the
disordered part of the chain show the expected bimodality but furthermore
also feature a peak centered around zero indicating the proximity of the thermal region.  
For the distribution of the thermal sites, we also notice the difference to the fix size setting discussed above that they do not
show a bimodality anymore but only heavier tails than in the case of low disorder.

To investigate these effects in a system size scaling, we again analyze
the standard deviation of the distributions in the two plots on the right of
Fig.\,\ref{full_l2}. Here, we find that the two regimes show the same qualitative behavior
namely a decaying standard deviation with increasing system size. As this is also the case
for both parts of the chain, it strongly suggests that the system tends towards narrower
distributions and hence delocalization upon increasing the system size.

Let us emphasize this last observation by showing the standard deviation of 
a site in the middle of the disordered part of the chain at $i =L-\lfloor L/4 \rfloor$
as a function of disorder strength in Fig.\,\ref{std_l2}.
We find that the gap between standard
deviations for different system sizes is diminished at higher disorder but
there is no crossover as in the case of a constant size thermal region.
For finite systems this implies that significantly stronger disorder scaling with the system size is needed in order to localize even the disordered part of the system. Furthermore, the extrapolation from the available system sizes leads to the conclusion that localization vanishes in the large system limit for this model as constants of motion become more non-local even in the regime of high disorder. 

Let us summarize the results. We find that constant thermal regions which
are independent of the system size seemingly can not hinder localization and their influence is suppressed by increasing the system size. 
An ergodic region that scales with the system size, however, changes the physics
of the model and apparently delocalizes the system. 
It is important to note that an ergodic bubble of the order of the system size is exponentially unlikely to appear. It
 is, however, unclear, if already a weaker scaling of the size of the thermal region with the system size would be sufficient to delocalize the system which is not reasonable to test with the system sizes available to exact diagonalization based schemes.

Let us relate our findings to results
obtained in Ref.\,\cite{luitz17} where the authors predict and numerically show
a parameter region in which a localized chain can be thermalized by a 
constant thermal bath. There the authors work in the 
effective description of MBL and present theoretical arguments for a possible
mechanism of instability of MBL phases. 
Note however, that the disorder strength in their model
is only implicitly defined. 
In Ref.\,\cite{luitz17} the authors use an effective model of the disordered part of the system in terms of quasi-local constants of motions. The disorder strength as used in our work here controls the locality of these constants of motion and by this changes the decay length of the coupling of the bath to different qLCOMs. Upon changing this decay length the authors of Ref.\,\cite{luitz17} identify a critical value whereupon a small thermal grain thermalizes the full system.
Based on the fact that we only find
such an instability for thermal regions which scale with the system size, we
present two explanations for this apparent contradiction. 

The first one being
that the two Hamiltonians do not exhibit similar physical signatures. 
Both display features of MBL, but the structure of the bath and coupling to the disordered part of the system might be incompatible. On the level of effective models and not the Hamiltonians as such, 
it may be that the two effective models in terms of qLCOMs
may be incompatible. In Ref.\,\cite{luitz17} the authors employ off-diagonal terms which
couple bath and MBL chain; we instead suggest that the important ingredient governing the localization
effects is solely the support of the qLCOMs and hence the diagonalizing unitary $U_D$. In the setting, 
in which the bath is constant, the unitary $U_D$ should still qualify to be 
quasi-local with an increased localization length
inside the bath, whereas for the extensive bath the unitary will not be local at all in sufficiently
large systems.

The second explanation is 
that the critical coupling decay length at the transition found 
in Ref.\,\cite{luitz17} might
actually correspond to the critical disorder strength separating the ergodic
and MBL phase, giving rise to the detected instability. 
Since no comprehensive
theory for this transition exists as well, it might be fruitful to combine
our findings with the arguments of Ref.\, \cite{luitz17} in order to possibly establish
a better understanding for the ergodic to MBL transition.

\section{Conclusions}

In this work, we have investigated the fate of the qLCOMs in the disordered
Heisenberg model 
in the presence of a small thermal region modelled by improbably low
disorder. 
We have examined the influence of these regions on the localization behavior. As such regions occur with high
probability in large systems this allows us insights into the stability of the MBL phase in the large system limit. 
As a measure of the locality of the qLCOMs we have employed
the expectation value of local magnetization operators obtained from
exact diagonalization.
In the system sizes accessible to us, we find that the qLCOMs and thereby the MBL phase is stable 
when coupled to a finite bath.
If the thermal region is allowed to scale with the system size, however,
our findings suggest that MBL will vanish when approaching the 
thermodynamic limit.

These results suggest that an isolated MBL system is stable upon increasing the system size. If, in contrast, coupled to an external bath, the question of the stability of MBL may depend on more subtle details as the precise size and shape of the bath may have a strong influence. An interesting further research direction would be to carry out a similar analysis for two dimensional systems and devise local probes for delocalization which could then be used in state of the art experimental realizations of MBL.

\section{Acknowledgements}
MG is grateful for the feedback on the manuscript by Nicolas
Tarantino as well as numerous discussion with the participants of the 
\emph{Anderson Localization and Interactions} workshop at the MPIPKS
-- specifically with Wojciech De Roeck, Alan
Morningstar, Anna Goremykina and Maksym Serbyn as well as Antonio Rubio Abadal
and Jun Rui  at MPQ. Moreover, we would like to
thank Henrik Wilming for many discussions
at earlier stages of this work. This work has been supported by the ERC (TAQ), the DFG (FOR 2724, CRC 183, EI 519/14-1), and the Templeton Foundation. This work has also received funding from the European Union's Horizon 2020
research and innovation programme under grant agreement No.~817482 (PASQuanS).

%
\end{document}